\documentclass[aps,twocolumn,pra,superscript,floatfix,superscriptaddress,showpacs,footinbib]{revtex4-1}

\usepackage{color,soul}
\usepackage{natbib}
\usepackage{amssymb}
\usepackage{eqnarray,amsmath}
\usepackage{graphicx}
\usepackage{dcolumn}
\usepackage{float}
\usepackage{color}
\usepackage{xcolor}
\usepackage[colorlinks=true,%
            linkcolor=blue,%
            urlcolor=blue,%
            citecolor=blue,%
            filecolor=blue,%
            bookmarksopen=true,%
            pdfauthor={F. G. Medina},%https://www.overleaf.com/10769193tdpqskjydmcs
            pdftitle={Fingerprints_MBSs_QR},%
            pdfsubject={Fingerprints_MBSs_QR_Manuscript},%
            pdfpagemode=UseOutlines]{hyperref}

\begin{document}

\date{\today}
%\title{Detection of Majorana bound states with tight-binding rings}
\title{Fingerprints of Majorana bound states in quantum-rings}
\author{F. G. Medina}
\author{J. P. Ramos-Andrade}
\author{L. Rosales}
\author{P. A. Orellana}
\affiliation{Departamento de F\'isica, Universidad T\'ecnica Federico Santa Mar\'ia, Casilla 110 V, Valpara\'iso, Chile}

\begin{abstract}

In this work, we investigate a quantum-ring coupled to a topological superconductor,  hosting Majorana bound states (MBSs). We study the MBSs effects over the spectrum and persistent current along the quantum-ring. To obtain physical quantities, we employ the  Green's function formalism. We found that the presence of MBSs leads to dramatic changes in the quantum-ring spectrum by inducing particle-hole symmetry. On the other hand, we have obtained a substantial reduction of the persistent current in the strong coupling limit between ring and MBSs depending on the ring parity. These behaviors provide a measurable signature of MBSs. Our findings could be used as additional information on MBSs presence in these quantum systems.

\end{abstract}

\maketitle

\section{Introduction}

Majorana fermions are particles which principal feature is to be their own antiparticles \cite{Majorana1937}. In condensed matter physics, they appear as zero-energy excitations (quasi-particles).
Due to that, the Majorana bound states (MBSs) satisfy non-Abelian statistics; they are of interest for quantum computation implementations\cite{alicea2016superconductors}. In this context, before going into the practical use of MBSs, it is relevant to reach its characterization to provide detection of them. Theoretically, a MBS is predicted to be found in a $p$-wave superconductor half-quantum vortex \cite{alicea2011non}. Several proposals have been carried out, to the detection of MBSs,  such as resonant Andreev reflection by a scanning tunneling microscope (STM) \cite{law2009majorana}, the $4\pi$ periodic Majorana-Josephson current \cite{fu2009josephson}, the implementation of optically trapped one-dimensional (1D) fermionic atoms \cite{jiang2011majorana} or an effective model in a cold-atom honeycomb lattice with textured pairings \cite{pan2018implementing}, among others. On the other hand, Kitaev developed a proposal based on a spinless chain proximitized by a superconductor.
This model represents  a  topological superconductor
at which MBSs appears localized at its ends \cite{kitaev2001unpaired}.

Recent experimental advances suggest that physical realization of Kitaev proposal has been carried out \cite{lutchyn2018majorana}.
In this scenario, zero-bias anomalies in transport quantities measured through TSCs have been seen as a way to detect MBSs presence \cite{Deng2012,mourik2012signatures,Lee2012,albrecht2016exponential,das2012zero,PhysRevB.87.241401}.
Since MBSs are exotic states, it is also interesting to explore its interplay with regular fermionic states, such as quantum dots (QDs).
For instance, the linear conductance through a QD with a side-coupled TSC exhibit a zero-bias distinctive characteristic;
 a half-integer conductance \cite{liu2011detecting,smirnov2015majorana,Vernek2014}. Interference phenomena in multiple QD-chains connected with MBSs have also been considered, proposing additional features for detection \cite{gong2014detection,zambrano2017majorana}. At this point, it is important to mention that the presence of TSCs hosting MBSs affected the transport phenomena, being a useful way for detection implementations.
Although theoretical results obtained in QD-MBS coupled systems cannot be refused, in physical realizations, resonant tunneling not always ensure the detection of MBSs mostly due to decoherence effects, which may suppress the signature of the MBSs. Therefore, it is attractive to study isolated systems and characterize any possible variations in their physical properties due to the presence of the MBSs.
An interesting isolated system is a quantum-ring crossed by a magnetic flux, also called $U\left(1\right)$ gauge field. Due to the break of the time-invariant symmetry, this system shows persistent currents, which corresponds to the generation of a spontaneous current as a response to changes in magnetic flux.  Persistent currents have been well studied in both experimentally and theoretically \cite{buttiker1983josephson,fuhrer2001energy,oh1995persistent,cheung1988persistent}, focusing on magnification \cite{orellana2005persistent}, controlling local currents \cite{yadalam2016controlling}, currents in 1D disordered rings \cite{chen2000persistent}, electron correlations \cite{chakraborty1994electron}, and even in alternative ways to generate persistent currents without a magnetic flux \cite{keck2018persistent}. Furthermore, persistent currents are sensitive to external perturbations, and under some circumstances, they present a robust behavior, which makes them good candidates to the detection of MBSs. In this sense, the use of persistent currents with the purpose of detection MBSs has been studied theoretically by Gong et al. \cite{gong2015odd}, where a TSC is embedded in the QD-ring. As a result, the relevant signature of the MBSs is the cancellation of persistent current signal when the ring parity is even. Ghazaryan et al. \cite{ghazaryan2016signatures} show that for a ring with a few hundred angstroms and a particular range of chemical potential, the system is in a topological phase with the possibility to find MBS.

In this work,  we study the electronic and transport properties of a quantum-ring side-coupled with a TSC hosting MBSs.  By using the Green's function formalism, we found a response in the quantum-ring spectra due to the connection with MBSs, which leads to establish a persistent current behavior depending on ring-TSC and/or MBSs coupling parameters. We believe our findings could be useful to provide further characterization of persistent currents in the presence of MBSs.
The manuscript is organized as follows: In  Sec. \ref{Sec.2} we present the model and theoretical background implemented along this work. In Sec. \ref{Sec.3}, we present the corresponding analytic results and their related discussions, and finally, we conclude by giving the final remarks in Sec. \ref{Sec.4}.

\section{Model} \label{Sec.2}

\begin{figure}[h]
   \centering
    \includegraphics[width=0.7\linewidth]{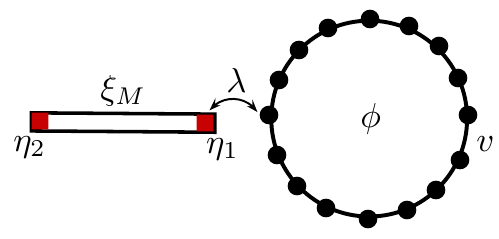}
    \caption{Schematic view of the model: TSC hosting MBSs at its ends (red) side attached to a QD-ring (connected black dots). The latter is crossed by a magnetic flux $\phi$.}
    \label{fig:RingMajorana2}
\end{figure}

The system under study consists of a discrete number of QDs arranged to form a quantum-ring coupled to a TSC, which is hosting MBSs at their ends, as is illustrated in Fig.\ \ref{fig:RingMajorana2}. We model the system by means of an effective low-energy Hamiltonian, which describes the electronic motion through the ring, in the presence of MBSs. The Hamiltonian is written as:

\begin{equation}
    \hat{H} = \hat{H}_{R} + \hat{H}_{M} + \hat{H}_{C}.
    \label{eq:Hamiltonian}
\end{equation}
where the term $\hat{H}_{R} = \sum_{l}v(\exp[i(2\pi/N)\Phi] c^{\dagger}_{l}c_{l+1} + \text{h.c.})$ corresponds to the quantum-ring Hamiltonian in the position space, where $v=v^{\ast}$ is the nearest-neighbor coupling parameter in the quantum-ring. Since the ring is crossed by a magnetic flux, a phase factor $\Phi = \phi/\phi_{0}$ is included, expressed in units of the quantum flux $\phi_{0} = h/e$. We are setting up the on-site energy at zero, and in order to perform the calculations, we express this Hamiltonian ($\hat{H}_{R}$) in the momentum space. Thus, we write
\begin{equation}
\hat{H}_{R} = \sum_{k} \varepsilon_kc^{\dagger}_{k}c_{k}\,,
\end{equation}
where $\varepsilon_{k}= 2v\cos\left[(2\pi/N)\Phi + ka \right]$.

The second term on the right-hand side of Eq.\ (\ref{eq:Hamiltonian}), $\hat{H}_{R}$, corresponds to the MBSs, which is described within the effective form described by Kitaev \cite{kitaev2001unpaired} as
\begin{equation}
  \hat{H}_{M} = i\xi_{M}\eta_{1}\eta_{2}\,,
\end{equation}
where $\eta_{\alpha}$ is the Majorana operator in the $\alpha$-edge ($\alpha = 1,2$) of the TSC. Majorana operators satisfy both $\{\eta_{\alpha},\eta^{\dagger}_{\beta}\} = \delta_{\alpha,\beta}$ and  $\eta_{\alpha} = \eta^{\dagger}_{\alpha}$, i.e. MBSs are characterized by being their own anti-particles. A useful way to treat them analytically is by means of regular fermionic operators superposition, as $\eta_{1} = \left(f + f^{\dagger}\right)/\sqrt{2}$ and $\eta_{2} =-i\left(f - f^{\dagger}\right)/\sqrt{2}$. The inter MBSs coupling $\xi_{M}$ describes the connection between both MBSs and it is proportional to $\xi_{M} \sim e^{-\ell/\ell_{0}}$, being $\ell$ the nanowire length and $\ell_{0}$ the superconducting coherence length. Note that in the case of $\xi_{M}=0$, both MBSs are strictly equivalent. The last term in the right hand side of Eq.\ (\ref{eq:Hamiltonian}) describes the coupling between MBSs and quantum-ring. We write it as
\begin{equation}
\hat{H}_{C} = \sum_{k}\lambda \left(c_{k}\eta_{1} + \eta_{1}c^{\dagger}_{k}\right)\,.
\end{equation}
Without loss of generality, we have chosen the coupling parameter between the ring and its nearest MBS to be real, i.e. $\lambda = \lambda^{\ast}$.

We will focus on the modifications of transport phenomena in the ring. The physical information about electronic transport is obtained employing the equilibrium Green's function techniques, in this case, we have implemented the equation of motion method, fully described in Ref. \cite{jishi2013feynman}. Therefore, the real-time retarded Green's function, that describes the motion of electrons in the quantum-ring is written as follows
\begin{equation}
    G^{r}\left(t\right) = -i\theta\left(t\right)\left<\left\{c_{k}\left(t\right),c^{\dagger}_{k}\left(0\right)\right\}\right>,
    \label{eq:retardedGreen}
\end{equation}
where $\left<\cdots\right>$ is the thermal average and $\theta\left(t\right)$ is the heaviside function, the anticommutator stands for the fermionic nature of the operators $c_k^\dagger \left(c_k\right)$.

The expression for the persistent current presents in the quantum-ring is obtained by

\begin{equation}
    I_{n} = -\frac{\partial E_{n}}{\partial \phi},
    \label{eq:persistentcurrent1}
\end{equation}
where $\phi$ is the magnetic flux, $E_{n}$ is the quantum-ring spectrum. The slope will determine if the persistent current is diamagnetic or paramagnetic. In the specific case in which we are considering spinless electrons, the persistent current is diamagnetic when the electron number is odd, and it is paramagnetic otherwise. This phenomenon is known as Leggett's conjecture \cite{leggett1991granular}. We obtain the total persistent current in the quantum-ring by summing over all $n$ states. \cite{cheung1989persistent}:
\begin{equation}
    I = - \underset{n}{\sum}f\left(\varepsilon\right)I_{n},
\end{equation}
where $f\left(\varepsilon\right)$ is the Fermi-Dirac distribution function.

\section{Results and discussions} \label{Sec.3}

In what follows, we will use $2v$ as the energy unit of the system. Based on the above description, the retarded Green's function for the system shown in Fig.\ \ref{fig:RingMajorana2} is obtained from Eq.\ (\ref{eq:retardedGreen}) in the energy domain. Then, the Green's function writes as

\begin{equation}
    G_{n}^{r}\left(\varepsilon\right) = \frac{1}{\varepsilon_{R} - \Sigma_{M}(\varepsilon,n)+i\epsilon},
    \label{eq:Green_funtionAB}
\end{equation}
where $\varepsilon_{R}=\varepsilon -\cos\left[(2\pi/N)\Phi + 2\pi n/N\right]$ and the self-energy $\Sigma_M\left(\varepsilon,n\right)$, which carries the information of the coupling between quantum-ring and TSC, is

\begin{align}
    \Sigma_M\left(\varepsilon,n\right) &= \\
    & \frac{ 2\varepsilon \Lambda \sum_{n'}g^{\left(-\right)}\left(\varepsilon,n'\right)}{g^{\left(-\right)}\left(\varepsilon,n\right)\left[ \varepsilon^{2} - \xi_{M}^{2} - 2\varepsilon\Lambda\sum_{n'}g^{\left(+\right)}\left(\varepsilon,n'\right) \right]},\nonumber \label{eq:self-energy}
\end{align}
where $g^{\left(\pm\right)}\left(\varepsilon,n'\right) = 1/\left(\varepsilon \pm \cos\left[(2\pi/N)\Phi + 2\pi n'/N\right]\right)$. Those are nothing but the isolated ring retarded Green's functions for holes and electrons respectively.
Given the self-energy, it is clear the way in which the MBSs are interacting effectively with the quantum-ring, since whenever $\Lambda = \left(\lambda/\sqrt{2}\right)^{2}\rightarrow 0 $, the self-energy becomes also zero, i.e. $\Sigma_M\left(\varepsilon,n\right)\rightarrow 0$, regardless other parameters in the system.

The spectrum for the Hamiltonian [Eq.\ (\ref{eq:Hamiltonian})] can be found from direct calculation, by means of recognizing the poles in  Green's function presented in Eq.\ (\ref{eq:Green_funtionAB}). What we expect to find is a polynomial type equation depending on the size of quantum-ring. Therefore, the poles are given by
\begin{equation}
    \frac{\varepsilon^{2} -2\varepsilon\Lambda\sum_{n'}\left[g^{\left(-\right)}\left(\varepsilon,n'\right) + g^{\left(+\right)}\left(\varepsilon,n'\right)\right] - \xi_{M}^{2}}{g^{\left(-\right)}\left(\varepsilon,n\right)} = 0\,,
    \label{eq:Polynomial}
\end{equation}
Note, the above expression has trivial zeros related to the system for $\Lambda = 0$, nothing but the isolated ring poles and the two MBSs of the TSC, by separate.

\begin{figure}[H]
    \centering
    \includegraphics[scale=0.53]{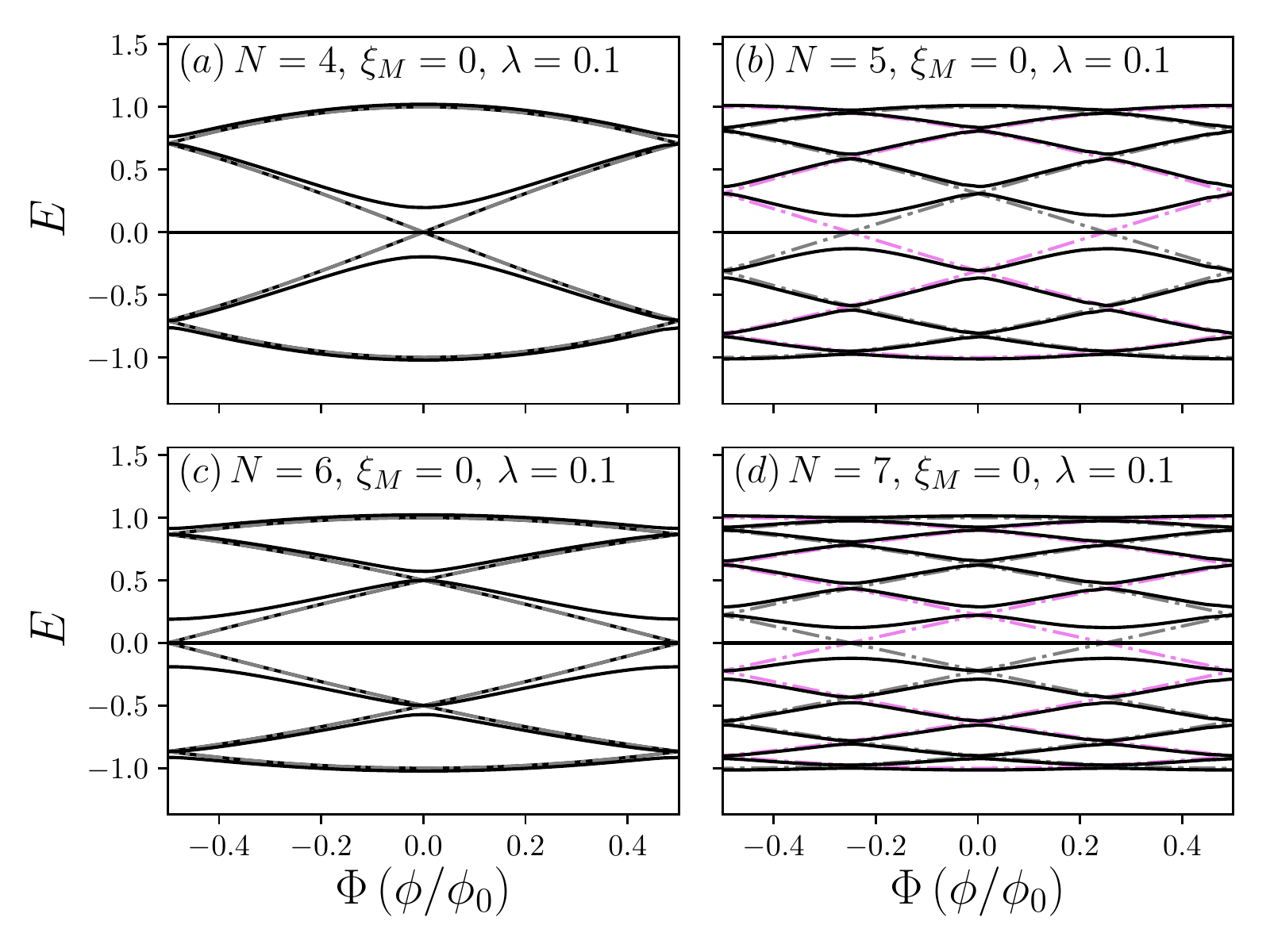}
    \caption{System spectra $E$ as function of magnetic flux $\Phi$, for fixed $\xi_{M} = 0$, using $\lambda = 0.1$ (black solid lines) and $\lambda=0$ (gray dashed-dotted line). [($a$) and ($c$)] $N$ even, and [($b$) and ($d$)] $N$ odd. In odd $N$ panels [($b$) and ($d$)], an electron-hole symmetric copy of the isolated ring spectrum is presented (violet dashed-dotted lines).}
    \label{fig:RMSpectrum(1)}
\end{figure}

Figure \ref{fig:RMSpectrum(1)} displays the quantum-ring spectra as a function of the dimensionless magnetic flux $\Phi$, for a given size $N$ of the quantum-ring. Solid black lines present the results for fixed $\lambda=0.1$ and the coupling between MBSs in the long superconducting wire limit, i.e., $\xi_{M} = 0$. Within this limit, the formation of a zero-energy eigenstate is observed, regardless both the magnetic flux and number of sites in the ring $N$.
The above is entirely due to the presence of a zero-energy MBS, a Majorana zero mode (MZM), in the system. The existence of this state can be understood from Eq.\ (\ref{eq:Polynomial}), by taking the case when $\xi_{M} = 0$, automatically we get $\varepsilon = 0$ as a solution, representing the pole at zero energy. Another relevant aspect is the number of states depending on even/odd $N$. For even $N$, the isolated ring spectra (gray dashed-dotted lines) fulfill electron-hole symmetry and crossing points are obtained at zero energy in $\Phi=0$ and $\Phi=\pm 0.5$ for $N=4$ and $N=6$, respectively. At this point, it is clear that the coupling of the MZM ($\lambda=0.1$) creates new states, holding the crossing and generating the anti-crossing points. The latter has a width of $2\lambda$. For odd $N$ ($N=5$ and $N=7$), the isolated quantum-ring spectrum does not fulfill electron-hole symmetry. Then, we obtain that whenever the MZM is connected, the spectra become electron-hole symmetric. Besides the zero energy state behaves as in even $N$ case, with a difference in the gap which is of width $\sqrt{2}\lambda$ and the crossing points are no longer observed. With the porpoise of understanding the spectra shape, we additionally plotted the negative spectrum of the isolated quantum-ring (violet dashed-dotted line). As a final remark, we go back to Eq.\ (\ref{eq:Polynomial}). We expect all solutions for $\varepsilon$ to be particle-hole symmetric. But, since $g^{\left(-\right)}\left(\varepsilon, n\right)$ with odd parity does not fulfill particle-hole symmetry, consequently these zeros are not allowed to be solutions. Regarding the even parity, these zeros fulfill particle-hole symmetry, and the spectra are gapless.

\begin{figure}[H]
    \centering
    \includegraphics[scale=0.53]{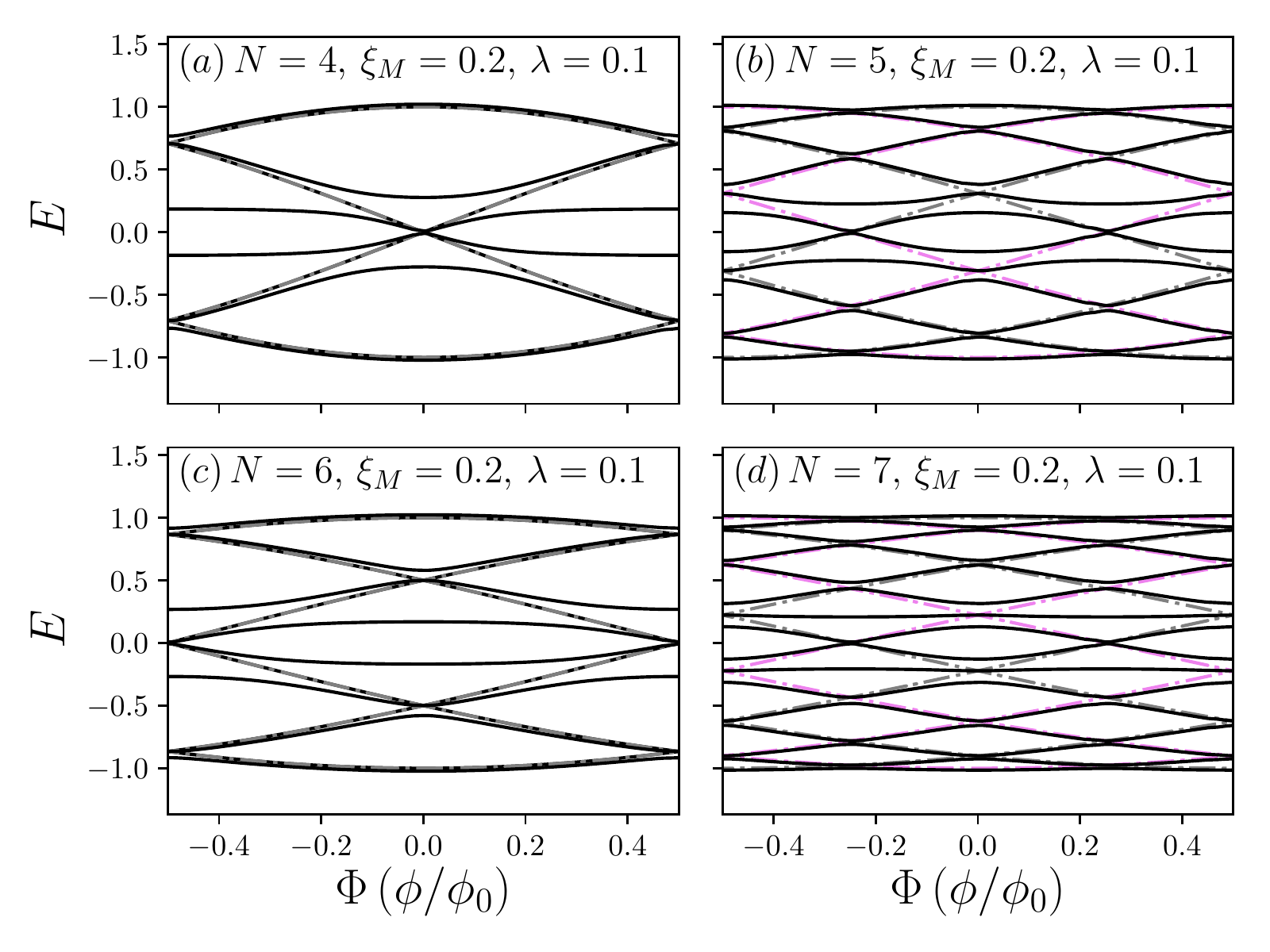}
    \caption{System spectra $E$ as function of magnetic flux $\Phi$, for fixed $\xi_{M} = 0.2$, using $\lambda = 0.1$ (black solid lines) and $\lambda=0$ (gray dashed-dotted line). [($a$) and ($c$)] $N$ even, and [($b$) and ($d$)] $N$ odd. In odd $N$ panels [($b$) and ($d$)], an electron-hole symmetric copy of the isolated ring spectrum is presented (violet dashed-dotted lines).}
    \label{fig:RMSpectrum(2)}
\end{figure}

\begin{figure}[H]
    \centering
    \includegraphics[scale=0.53]{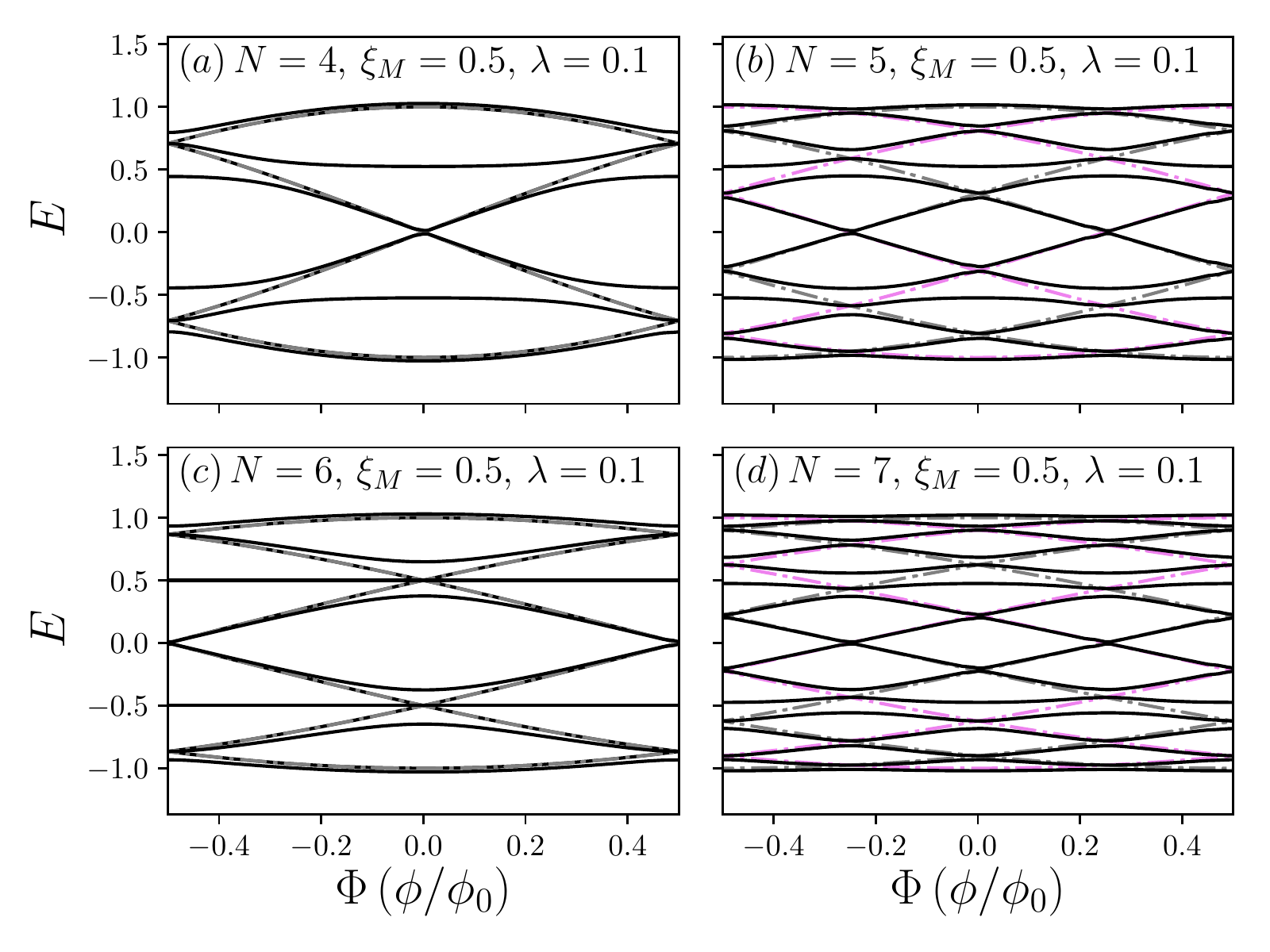}
    \caption{System spectra $E$ as function of magnetic flux $\Phi$, for fixed $\xi_{M} = 0.5$, using $\lambda = 0.1$. [($a$) and ($c$)] $N$ even, and [($b$) and ($d$)] $N$ odd.}

    \label{fig:RMSpectrum(3)}
\end{figure}

In Fig.\ \ref{fig:RMSpectrum(2)} the case with $\xi_{M}=0.2\neq 0$ is addressed. For even $N$, the degeneracy of the MZM is removed, obtaining states at $\pm\xi_{M}$, except around the crossing points at zero energy, i.e., around $\Phi=0$ for $N=4$ and $\Phi=\pm 0.5$ for $N=6$. For odd $N$, crossing points at zero energy are now observed placed at $\Phi=\pm 0.25$, for $N=5$ and $N=7$. It is remarkable how the spectra become electron-hole symmetric whenever the MBSs are connected, regardless of $N$. Also, it is worth to mention that the number of states increases from $N$ (isolated ring) to $2N+2$ when the MBSs are coupled.

Figure\ \ref{fig:RMSpectrum(3)} displays the case with strong coupling between MBSs, $\xi_{M}=0.5$. It is evident that the central energy range, around zero energy, becomes wider as the value of $\xi_{M}$ increases. The latter becomes relevant since the splitting of the states takes place at the edges of the central energy range;  i.e., tuning this, it is possible to control the energy where the splitting of the state occurs. Whenever $\xi_{M}\neq 0$, four-level crossing and anti-crossings are observed around zero energy. The involved states here are the ones corresponding to MBSs and those related to isolated quantum-ring spectrum (gray dashed-dotted lines). In the case of even $N$, the anti-crossing are similar to what was described above, as was described above.  While for odd $N$, crossings are also observed between MBSs and related to electron-hole symmetric isolated ring spectrum (dashed-dotted gray and violet lines).

A further analysis for the spectra central energy range is performed at the limit $\xi^{2}_{M} >> \Lambda$, the self energy is reduced to a more simple form, given by $\Sigma_{M} \sim - 2\varepsilon\Lambda\sum_{n'}g^{-}\left(\varepsilon,n'\right)/\xi^{2}_{M}g^{\left(-\right)}\left(\varepsilon,n'\right)$, and consequently, the Green's function becomes

\begin{equation}
G^{R}\left(\varepsilon\right) \sim  \left[\varepsilon_{R} + \frac{2\varepsilon \Lambda \sum_{n'} g^{\left(-\right)}\left(\varepsilon,n'\right)}{\xi^{2}_{M} g^{\left(-\right)}\left(\varepsilon,n'\right)}\right]^{-1}. \label{eq:Green's_function2}
\end{equation}
 Therefore, the two limits in which the Green's function returns to the isolated quantum-ring case are obtained. The first, and trivial one, occurs when $\Lambda\rightarrow0$, and the second when $\xi^{2}_{M} >> \Lambda$. Both seem completely equivalent in the ratio of Eq.\ (\ref{eq:Green's_function2}), but physically corresponds to two different cases. The trivial limit corresponds to the quantum-ring without any TSC attached to it. The second one when the MBSs are strongly coupled, and the TSC is still connected with the quantum-ring, i.e., MBSs form a fermionic state and particle-hole symmetry is broken.

A subtle difference will arise in the persistent currents between these two limits, since both reduce to the behavior of an isolated quantum-ring. Hence, when $\lambda$ takes large values, splitting in the bands produces less participation of these states in the conduction of electrons in the quantum-ring, and new states are now visible for finite values of $\xi_{M}$. From here, it is clear the effect of the MBSs observing $\Sigma_{M}$.

\begin{figure}[H]
    \centering
    \includegraphics[scale=0.62]{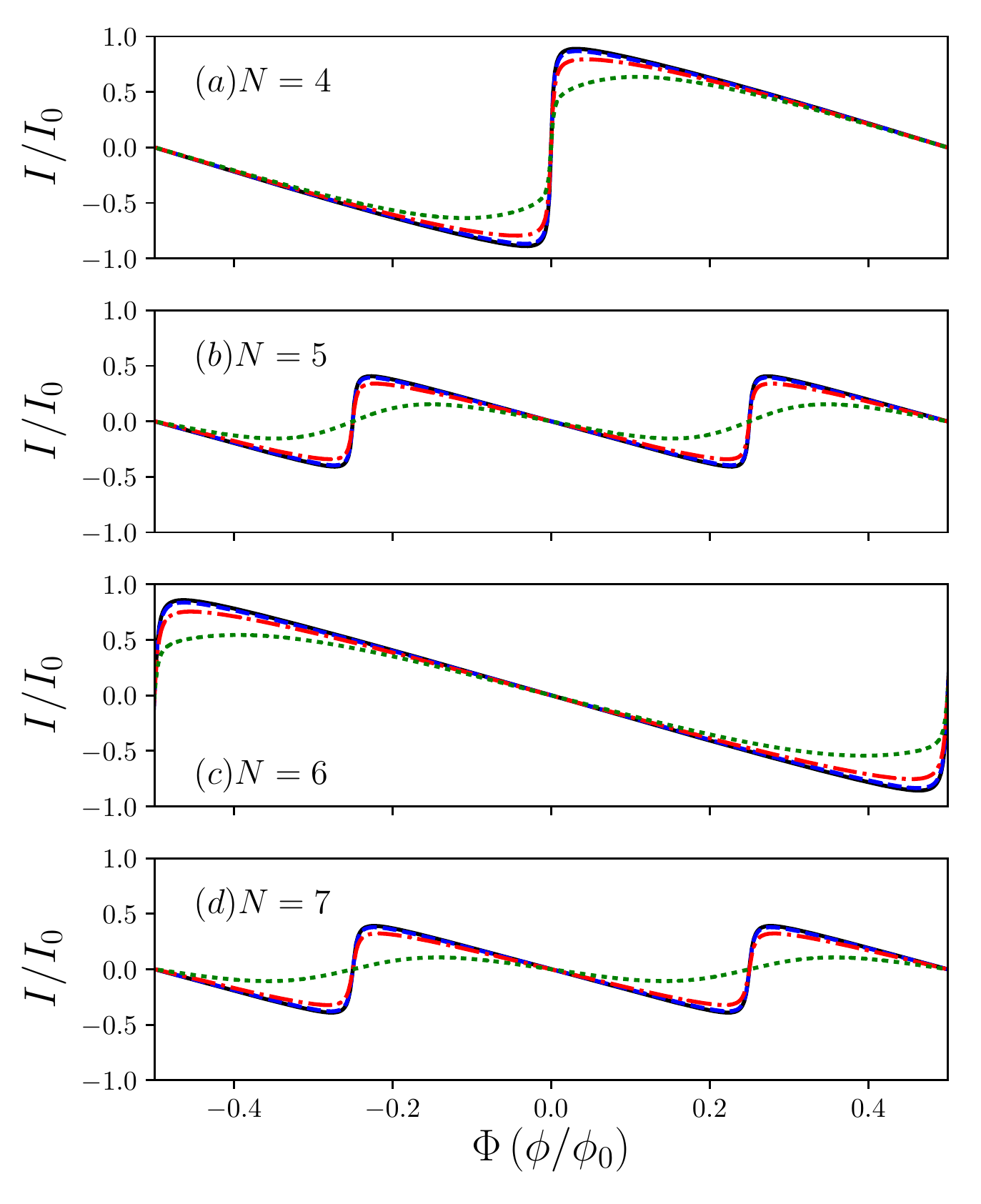}

    \caption{Dimensionless persistent current $I/I_{0}$, where $I_{0} = 2ve/\hbar N$, as a function of the dimensionless magnetic flux $\Phi$, for fixed $\lambda = 0.1$. ($a$) $N = 4$, ($b$) $N = 5$, ($c$) $N = 6$, and ($d$) $N = 7$. In all panels, $\xi_{M}$ takes values $\xi_{M} = 1$ (black solid line), $\xi_{M} = 0.5$ (blue dashed line), $\xi_{M} = 0.2$ (red dashed-dotted line), and $\xi_{M} = 0$ (green dotted line).}
    \label{fig:RMCurrent}
\end{figure}

\begin{figure}[h]
\centering
    \includegraphics[scale=0.7]{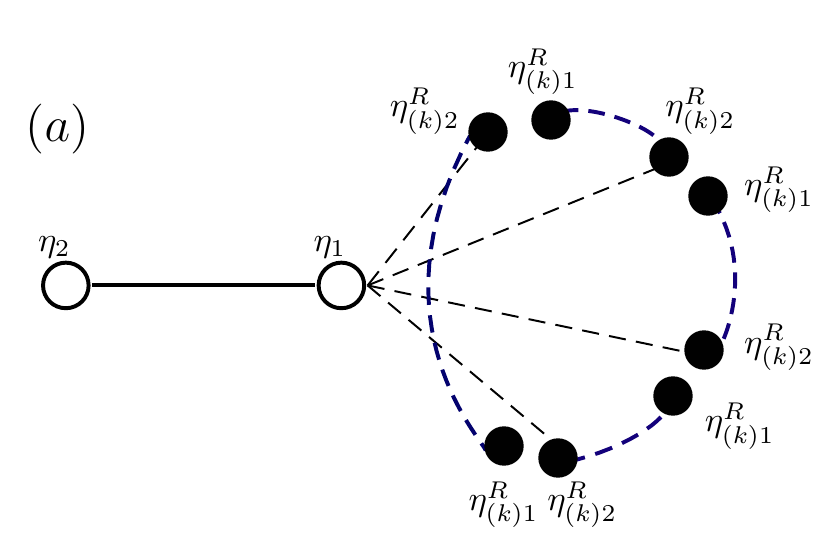}
    \includegraphics[scale=0.7]{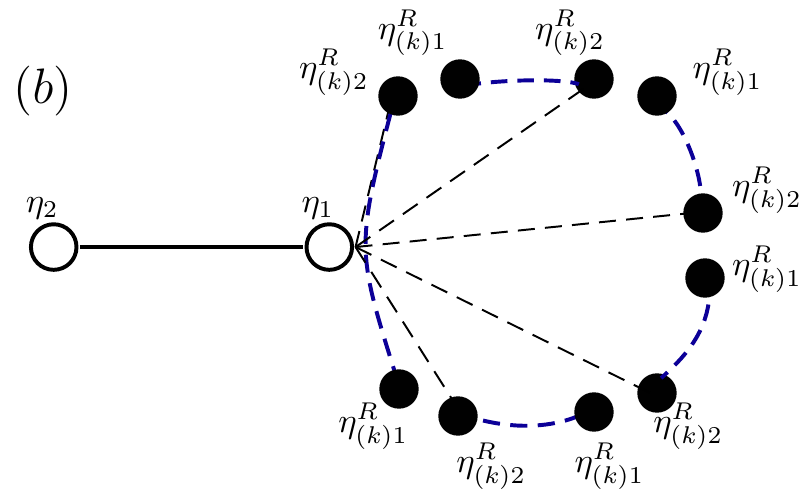}\\
\caption{Majorana representation of the ring in momentum space. Blue dashed line describes the coupling between ring states, while black dashed lines are the coupling between TSC and the MBSs $\eta^{R}_{\left(k\right)2}$ and the solid line are the representation of TPS hosting MBSs at the edges. $(a)$ representation for $N = 4$, and $(b)$ representation for $N = 5$.}
\label{fig:RM_representation}
\end{figure}

All the features of the spectra discussed above have an impact on persistent currents behavior, as we will discuss in what follows. Figure \ref{fig:RMCurrent} displays the persistent currents obtained for different values of $\xi_{M}$, for different values of even/odd $N$, from $N=4$ to $N=7$, from top to bottom panels, respectively. These results were obtained within zero-temperature limit, $T=0$, and for fixed $\lambda = 0.1$. In this case, for a given $\lambda$, in all panels, the current amplitude is maximum when $\xi_{M} = 1$ (solid lines), and decrease continuously as $\xi_{M}$ decreases (non-solid lines). Another relevant aspect to note is the persistent current shape. A well-formed saw-tooth shape is obtained for $\xi_{M}=1$ (solid black lines), being this the characteristic of a persistent current of an isolated ring. For $\xi_{M} = 0.5$ (blue dashed lines), a softer saw-tooth shape is still present, while for $\xi_{M} = 0.2$ (red dashed-dotted lines) the saw-tooth looses its form and starts looking like a sinusoidal function for odd $N$. Finally, with $\xi_{M} = 0$ (green dotted lines) the persistent current is given by a $I \propto -\sin\left(\Phi\right)$ function whatever the odd $N$ taken.

On the other hand, for even $N$, the persistent current changes iteratively. This behavior is due to phase changes, as it is observed in the spectra. Besides, the effects on the persistent current amplitude due to the coupling between MBSs is different in comparison with Ref.\ \cite{gong2015odd}. It is due to the distinction in the configuration of the system. While in Ref.\ \cite{gong2015odd}, the TSC is embedded in the quantum-ring, in our case we consider the interaction between TSC and the quantum-ring through proximity effect. It is fundamental since for a TSC embedded in a quantum-ring brakes down the periodicity of a ring.

In order to obtain a complementary notion of how MBSs couples to the quantum-ring, we  write out the full Hamiltonian [Eq.\ (\ref{eq:Hamiltonian})] in a Majorana representation, i.e. expressing the fermionic operators belonging to the ring in terms of Majorana operators. Henceforth, generic fermionic operators $f$ and $f^{\dagger}$ are written in the form $f\sqrt{2} = \eta_{1} + i\eta_{2}$ and $f^{\dagger}\sqrt{2} = \eta_{1} - i\eta_{2}$, respectively. The dimensionless Hamiltonian of the ring is then written as $\hat{H}_{R} = \sum_{k}\left(1/2 + i\eta^{R}_{\left(k\right)1}\eta^{R}_{\left(k\right)2}\right)\cos\left[\left(2\pi/N\right)\Phi + ka\right]$. The coupling Hamiltonian between the ring and the TSC changes to $\hat{H}_{C} = \sum_{k}i\lambda\sqrt{2}\eta^{R}_{\left(k\right)2}\eta_{1}$. Lastly, the TSC Hamiltonian just remains as $\hat{H}_{M} = i\xi_{M}\eta_{1}\eta_{2}$. An schematic view of this transformation is presented in Fig.\ \ref{fig:RM_representation}, for $N=4$ and $N=5$.

%\begin{widetext}
\begin{figure*}[tbph]
    \centering
    \includegraphics[width=\textwidth]{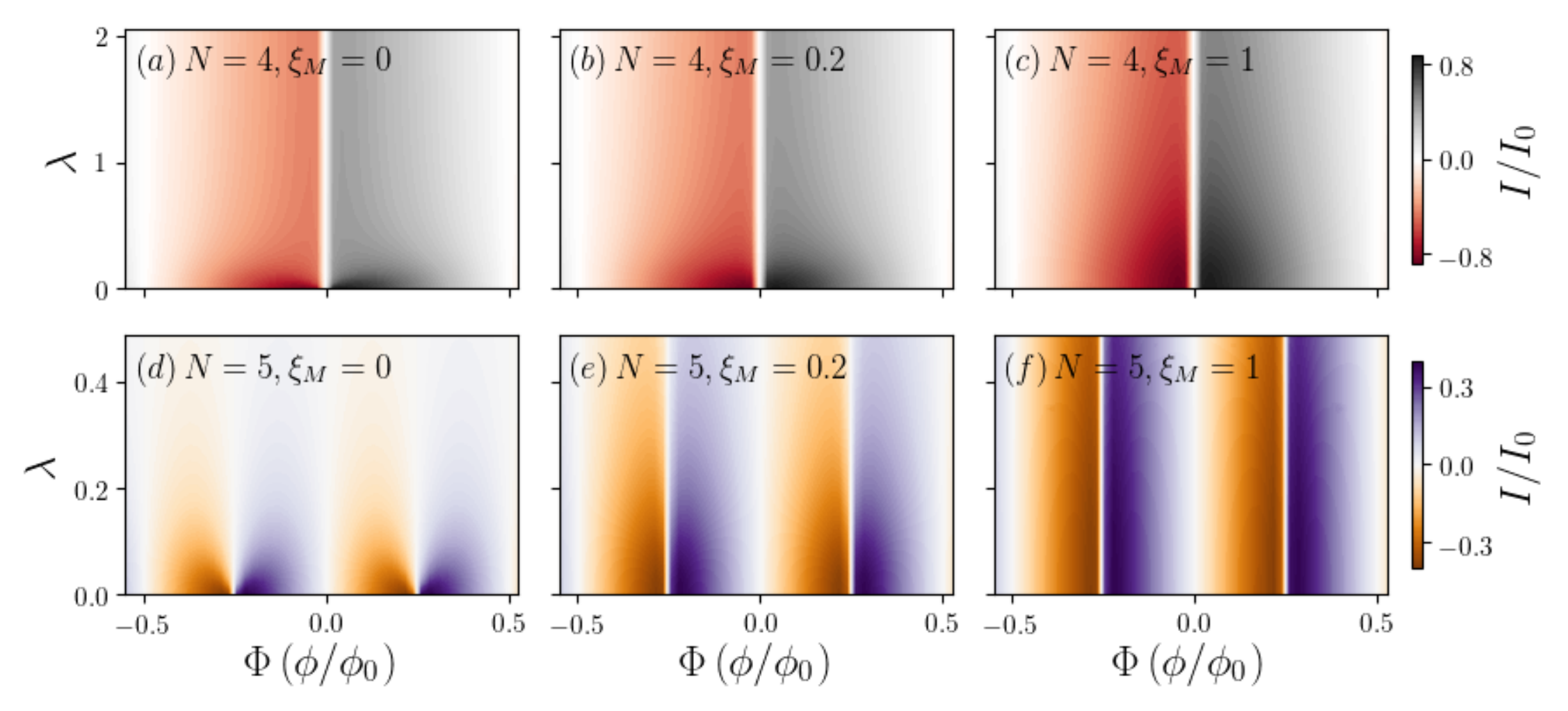}
    \caption{Dimensionless persistent current $I/I_{0}$, where $I_{0} = 2ve/\hbar N$, as a function of the dimensionless magnetic flux $\Phi$ and the coupling with the TSC $\lambda$. We used $\xi_{M} = 0$ [($a$) and ($d$)], $\xi_{M} = 0.2$ [($b$) and ($e$)], and $\xi_{M} = 1$ [($c$) and ($f$)]. $N = 4$ is used in upper panels and $N=5$ in lower panels.}
    \label{fig:Contour1}
\end{figure*}
%\end{widetext}

The above representation allows us to interpret a non-vanishing coupling between all states in the system. Accordingly, $\hat{H}_{R}$ in Majorana representation shows that the quantum-ring is mapped into a Majorana quantum-ring.  Here, there is a connection between MBSs with opposite helicity, with the preservation of the periodic boundary conditions. It explains why for odd and even $N$, we achieved not much change in the persistent currents with the weak presence of the MBSs. Finally, Majorana representation of $\hat{H}_{C}$ shows that $\eta_{1}$ couples to all $\eta^{R}_{\left(k\right)2}$.

Figure\ \ref{fig:Contour1} shows the persistent current as function of $\lambda$ and $\Phi$, for different values of $\xi_{M}$ and for $N=4$ (5) in upper (lower) panels. The case of $\xi_{M} = 0$ is addressed in Figs.\ \ref{fig:Contour1}($a$) and \ref{fig:Contour1}($b$), from which it is observed a non vanishing persistent current, as long as $\hat{H}_{C}$ is less relevant than $\hat{H}_{R}$ to the conduction of electrons. In other words, whenever the TSC is strongly coupled to the ring, the persistent current goes to zero, but we have to be careful, since we show before than in the even parity particle-hole symmetry allows the zeros $g^{\left(-\right)}\left(\varepsilon,n\right)$. Therefore, the persistent current does not cancel out in even parity, and we can observe a quench in the persistent current signal, contrary to what happened in the odd parity when the zeros $g^{\left(-\right)}\left(\varepsilon,n\right)$ are not allowed.
In the strong coupling case, the blue dashed lines disappear and the edge state of the TSC couples strongly with the quantum-ring states, i.e., $\eta^{R}_{\left(k\right)1}$ are decoupled to the $\eta^{R}_{\left(k\right)2}$. Hence, as a consequence the circuit is opened, and the persistent current goes to zero, this effect has a counterpart in the even parity, since isolated ring states are somehow protected by the particle-hole symmetry, no matter how strong the coupling MZM and quantum-ring is, always we can find a tiny signal in the persistent current.

For  $\xi_{M} \neq 0$ and weak coupling ring-TSC ($\lambda \sim 0$)  the persistent current it is perfectly observed as a saw-tooth signal in all panels of Fig.\ \ref{fig:Contour1}. By increasing $\lambda$, it is obtained a strong quenching in the persistent current amplitude for $\xi_{M}=0.2$ [Figs.\ \ref{fig:Contour1}($b$) and \ref{fig:Contour1}($e$)] and a subtle quenching for $\xi_{M} = 1$ [Figs.\ \ref{fig:Contour1}($c$) and \ref{fig:Contour1}($f$)], both in contrast to the persistent current vanishing obtained for $\xi_{M}=0$. In order to understand this behaviour in the persistent current, we revisit the spectra for values $\xi_{M} \neq 0$ showed in Fig.\ \ref{fig:RMSpectrum(2)} and Fig.\ \ref{fig:RMSpectrum(3)}. From these, we can observe the splitting in the ring states due to TSC connection. However due to the split MBSs, appears branches replicating the original states of the isolated ring, but also fulfilling particle-hole symmetry as was discussed above. Furthermore, the persistent currents do not cancel out due to the formation of those central branches. These states, even with large $\lambda$, generate persistent currents with a more relevant contribution in the amplitude for larger values of $\xi_{M}$. In more specific words, the particle-hole symmetry is preserved, and to satisfy this condition, when $\xi_{M} \neq 0$ the symmetric branch around zero energy appear. On the other hand, for $\xi_{M} > 1$, both MBSs present in the TSC form a fermionic state, and consequently, the particle-hole symmetry is no longer fulfilled. Therefore, within this limit, the persistent current has a tendency to keep its amplitude, as in Fig.\ \ref{fig:Contour1}($c$) and Fig.\ \ref{fig:Contour1}($f$).

\section{Summary} \label{Sec.4}

In the present work, we study the spectrum and persistent current of a quantum-ring side-coupled to a TSC supporting MBSs. By employing the Green's function formalism using the equation of motion method, we investigate effects due to the leakage of MBSs in spectra and persistent currents. Spectra analysis is carried out on the weak coupling regime for the interaction between TSC and quantum-ring ($\lambda = 0.1$).
We analize three cases for the coupling between the MBSs: the limit with long TSC wire ($\xi_{M} = 0$), a weak coupling case ($\xi_{M} = 0.2$), and when the coupling is $\xi_{M} = 0.5$.
In all cases, even and odd $N$ has been considered. For $\xi_{M} = 0$, a robust zero energy state produce an anti-crossing at the degeneracy points for even $N$.
The other states that appear in the even parity behave as the isolated quantum-ring states and those remains intact no matter how strong the coupling is.  It is understood since particle-hole symmetry allows those states to exist in even parity.
For the case of odd parity, the anti-crossing also takes place with the isolated quantum-ring states, the MZM and the electron-hole symmetric copy of the isolated quantum-ring states. The latter one can be understood due to the induced particle-hole symmetry in the ring, which is preserved for finite values of $\xi_{M}$. In the case when $\Sigma_{M}\left(\varepsilon,n\right) \rightarrow 0$ particle-hole symmetry is broken, since in this limit MBSs form a fermionic state. On the other hand, we show that the presence of a MZM produces a strong quench in the persistent currents for even $N$ and cancels out the persistent current for odd $N$ in the quantum-ring, whenever the TSC is strongly coupled to the quantum-ring. We believe our findings could be useful to establish additional measurable signatures of MBSs presences.

\begin{acknowledgments}

F.G.M. is grateful for the funding of scholarship CONICYT-Chile No. 21170550. J.P.R.-A is grateful for the funding of FONDECYT Postdoc. Grant No. 3190301 (2019). L.R. and P.A.O. acknowledges support from FONDECYT Grant No. 1180914.

\end{acknowledgments}

\bibliographystyle{apsrev4-1}
\bibliography{biblio}

\end{document}